\documentclass[12pt,preprint]{aastex}

\newcommand{\ls}{\object{LS~5039}}
\newcommand{\lsf}{\object{LS~5039}/\object{RX~J1826.2$-$1450}}
\newcommand{\lsff}{LS~5039/RX~J1826.2$-$1450}
\newcommand{\eg}{\object{3EG~J1824$-$1514}}

\slugcomment{Accepted for publication in ApJ}

\begin{document}

\title{ORBITAL X-RAY VARIABILITY OF THE MICROQUASAR \ls}

\author{Valent\'{\i} Bosch-Ramon\altaffilmark{1},
Josep M. Paredes\altaffilmark{1},
Marc Rib\'o\altaffilmark{2},
Jon M. Miller\altaffilmark{3,4},
Pablo Reig\altaffilmark{5,6},
and Josep Mart\'{\i}\altaffilmark{7}}

\altaffiltext{1}{Departament d'Astronomia i Meteorologia, Universitat de
Barcelona, Av. Diagonal 647, 08028 Barcelona, Spain; vbosch@am.ub.es;
jmparedes@ub.edu}
\altaffiltext{2}{Service d'Astrophysique and AIM, CEA Saclay, B\^at. 709, L'Orme des
Merisiers, 91191 Gif-sur-Yvette Cedex, France; mribo@discovery.saclay.cea.fr}
\altaffiltext{3}{Harvard-Smithsonian Center for Astrophysics, 60 Garden
Street, Cambridge, MA02138, USA; jmmiller@head-cfa.harvard.edu.}
\altaffiltext{4}{National Science Foundation Astronomy and Astrophysics
Postdoctoral Fellow, USA}
\altaffiltext{5}{GACE, Instituto de Ciencias de los Materiales, Universitat de
Val\`encia, 46071 Paterna-Val\`encia, Spain; pablo.reig@uv.es}
\altaffiltext{6}{IESL (FORTH) \& Physics Dept. (University of Crete), 
P.O.Box 208, 71003, Heraklion, Greece; pau@physics.uoc.gr}
\altaffiltext{7}{Departamento de F\'{\i}sica, Escuela Polit\'ecnica Superior,
Universidad de Ja\'en, Virgen de la Cabeza 2, 23071 Ja\'en, Spain; jmarti@ujaen.es}

\shorttitle{Orbital X-ray variability of LS~5039}

\shortauthors{Bosch-Ramon et~al.}

\begin{abstract}
The properties of the orbit and the donor star in the high mass X-ray binary
microquasar \ls\ indicate that accretion processes should mainly occur via a
radiatively driven wind. In such a scenario, significant X-ray variability
would be expected due to the eccentricity of the orbit. The source has been
observed at X-rays by several missions, although with a poor coverage that
prevents to reach any conclusion about orbital variability. Therefore, we
conducted {\it RossiXTE} observations of the microquasar system \ls\ covering
a full orbital period of 4 days. Individual observations are well fitted with
an absorbed power-law plus a Gaussian at 6.7~keV, to account for iron line
emission that is probably a diffuse background feature. In addition, we have
taken into account that the continuum is also affected by significant diffuse
background contamination. Our results show moderate power-law flux variations
on timescales of days, as well as the presence of miniflares on shorter
timescales. The new orbital ephemeris of the system recently obtained by
Casares et~al. have allowed us to show, for the first time, that an increase
of emission is seen close to the periastron passage, as expected in an
accretion scenario. Moreover, the detected orbital variability is a factor of
$\sim$4 smaller than the one expected by using a simple wind accretion model,
and we suggest that an accretion disk around the compact object could be
responsible for this discrepancy. On the other hand, significant changes in
the photon index are also observed clearly anti-correlated with the flux
variations. We interpret the overall X-ray spectral characteristics of \ls\ in
the context of X-ray radiation produced by inverse Compton and/or synchrotron
processes in the jet of this microquasar.
\end{abstract}


\section{Introduction}

\lsf\ is a high-mass X-ray binary runaway microquasar at a distance of
$\sim$2.9~kpc from the Earth \cite{paredes00,ribo02}. Paredes et~al. (2000)
proposed that this microquasar is the counterpart of the high-energy
$\gamma$-ray source \eg\ \citep{hartman99}, and suggested that microquasars
could be the counterparts of some of the unidentified EGRET sources. In
addition, Torres et~al. (2001) classified this source as variable, and
Bosch-Ramon \& Paredes (2004) have proposed a model based on an inhomogeneous
jet, emitting at $\gamma$-rays via inverse Compton (IC) scattering, giving
theoretical support to the possible association between the microquasar and
the EGRET source. 

The first radio detection was reported by Mart\'{\i}, Paredes, \& Rib\'o
(1998) using the VLA, but the discovery of the radio jets was only possible
when the source was observed at milli-arcsecond scales with the VLBA
\citep{paredes00}. The presence of an asymmetric and persistent radio jet has
been confirmed with new MERLIN and EVN observations \citep{paredes02}. The
radio emission is persistent, non-thermal and variable, although no strong
radio outbursts or periodic variability have been detected so far (Rib\'o
et~al. 1999; Rib\'o 2002).

In the optical band, \ls\ appears as a bright $V$=11.2, O6.5V((f)) star
showing little variability on timescales of months to years \citep{clark01}.
Variations of $\sim$0.4~mag have been reported in the infrared ($H$ and $K$
bands) but no obvious mechanisms for such variability have been proposed
\citep{clark01}. Using the radial velocities reported in McSwain et~al. (2001,
2004) and newly obtained ones, Casares et~al. (2005) have determined the most
recent values for the orbital period and the eccentricity for this binary
system, being $P_{\rm orb}=3.9060 \pm 0.0002$~d (with $t_0={\rm
HJD}~2$,451,$943.09 \pm 0.10$) and $e=0.35 \pm 0.04$, respectively, with the
periastron passage at phase 0.0.

In the X-ray domain the source has been observed several times and using
different satellites, presenting fluxes spanning one order of magnitude in the
range $\sim 5$--$50\times 10^{-12}$ erg~cm$^{-2}$~s$^{-1}$ (see
Sect.~\ref{xrays}). These fluxes imply luminosities in the range $\sim
0.5$--$5\times 10^{34}$~erg~s$^{-1}$, assuming a distance of 2.9~kpc.
Variations in the photon index have also been observed. So far, it has been
unclear whether these variations in flux and spectral slope were related to
long-term variations in the mass-loss rate of the companion or to accretion
changes along the eccentric orbit. 

To determine the possible flux and photon index variations along a whole
orbital period, we proposed X-ray observations of \ls\ to be performed with
the {\it RXTE} satellite during 4 consecutive days in 2003 July. Photometric
and spectroscopic optical observations were performed simultaneously to the
X-ray ones. Regarding the photometry, Mart\'{\i} et~al. (2004) have reported
that no photometric variability above $\pm0.01$~mag took place during six
consecutive nights. Regarding the spectroscopy, Casares et~al. (2005) have
found a nearly constant equivalent width of the H$\alpha$ absorption line of
$EW\sim 2.8$~\AA, implying no significant changes in the mass-loss rate of the
primary during our {\it RXTE} observations.

This work has been organised as follows. In Sect.~\ref{xrays}, we present
results from previous X-ray observations. In Sect.~\ref{rxte}, we show the
results obtained from our {\it RXTE} data reduction and analysis. Further, in
Sect.~\ref{disc}, we discuss our results in the context of both previous
observations and the microquasar scenario and finally, in Sect.~\ref{sum}, we
summarise this work.

\section{Previous X-ray results} \label{xrays}

The first X-ray detection of \ls/RX~J1826.2$-$1450 was made with {\it ROSAT}
in 1996 October, obtaining a flux of $7.1\times 10^{-12}$
erg~cm$^{-2}$~s$^{-1}$ in the range 0.1--2.4~keV \citep{motch97}. Rib\'o
et~al. (1999) reported {\it RXTE} observations carried out in 1998 February 8
and 16. The flux detected in the energy range 3--30~keV was $\sim 50\times
10^{-12}$ erg~cm$^{-2}$~s$^{-1}$, and the obtained photon index was
$\Gamma=$1.95, although these values are likely affected by contamination from
the diffuse background (see discussion below). The X-ray timing analysis
indicated the absence of pulsed or periodic emission on timescales of
0.02--2000 seconds. The source spectrum was well represented by a power-law
model plus a Gaussian component describing a strong iron line at 6.7~keV.
Significant emission was seen up to 30~keV, and no exponential cut-off at high
energy was required. Unpublished observations performed by {\it ASCA} (1999
October 4) present a photon index of about 1.6, and an unabsorbed flux in the
range 0.3--10~keV of about $13\times 10^{-12}$~erg~cm$^{-2}$~s$^{-1}$
(Martocchia, Motch, \& Negueruela 2005). {\it BeppoSAX} observations, on 2000
October 8, found a flux of $\sim 5\times 10^{-12}$ erg~cm$^{-2}$~s$^{-1}$
(0.3--10 keV), a photon index of 1.8 and a hydrogen column density of $N_{\rm
H}=1.0^{+0.4}_{-0.3}$~cm$^{-2}$ \citep{reig03a}. These authors also concluded
that an X-ray eclipse by the companion star was ruled out by these data. The
new ephemeris obtained by Casares et~al. (2005) confirm that the {\it
BeppoSAX} observations took place during the inferior conjunction of the
primary. The lack of eclipses provides an upper limit for the inclination of
the system of $i<69\pm1\degr$ (see discussion in Reig et~al. 2003a for
details). We have analysed unpublished {\it Chandra} data (2002 September 10)
and have found a flux of $8.9\times 10^{-12}$ erg~cm$^{-2}$~s$^{-1}$ and
$\Gamma=$1.14$\pm$0.12 in the range 0.3--10~keV. Finally, {\it XMM}
observations of \ls\ were carried out recently (2003 March 8 and 27), and
showed that the source presented a flux of about $10\times10^{-12}$
erg~cm$^{-2}$~s$^{-1}$ (0.3--10~keV) and a photon index of about 1.5
(Martocchia et~al. 2005). very similar fluxes and photon indexes are found for
both observing periods, being consistent with the fact that both were taken at
the same orbital phase (around 0.55 with the new ephemeris).

In any but the {\it RXTE} observations in 1998, a Gaussian component in
addition to the absorbed power-law model does not improve the fit. Therefore,
it is possible that the Gaussian line of the 1998 {\it RXTE} observations
could come from the Galactic ridge, being present because of the large field
of view and effective area of the instrument. We investigate this issue below.
For all the observations, disk features are not necessary to fit the data.

We list in Table~\ref{priorobs} the dates, missions, orbital phases, photon
indexes and extrapolated fluxes in the energy range 3--30~keV of \ls\ using a
power-law spectrum, assuming the same slope all up to 30~keV, for all the
prior observations but the {\it ROSAT} one (because of the big difference in
energy range). Although it is clear that photon index and flux changes are
present, since the different observations are separated months/years in time,
it has not been possible up to now to clearly establish whether the intensity
of the emission changed with time due to variations in the mass-loss rate of
the stellar companion or due to the orbital eccentricity. There is
observational evidence supporting the first option, based on correlations
found between the wind intensity of the companion (inferred from the
equivalent width of the H$\alpha$ absorption line) and the observed X-ray flux
(Reig et~al. 2003a; McSwain et~al. 2004), although this does not preclude the
second one to occur as well. The photon index variation could also be a
consequence of either long-term phenomena, short-term phenomena, or both.

\section{{\it RXTE} observations} \label{rxte}

The main goal of these observations was to perform an in depth study of the
X-ray variability during a full orbital period of the microquasar \ls. In
particular, we wanted to answer the following questions: Does the source flux
vary with orbital phase due to a geometric effect related to the eccentricity
of the binary system? Is the power-law index dependent on the orbital phase?
Does the iron line come from the source? This section tries to answer these
three queries, although a deeper and wider discussion is developed in
Sect.~\ref{disc}.

\subsection{Data reduction and analysis}

{\it RXTE} observed the source on 17 runs of different duration covering a
total of 4 days between 2003 July 4 and 8. Due to the relatively faint
emission of \ls\ during the observations, the source spectrum is background
dominated beyond 30~keV, so HEXTE data has not been considered and we have
focused on the analysis of PCA data. For the four longer runs, only PCU
\#\,2\footnote{Concerning PCU \#\,0, despite efforts to recalibrate its
response function following the loss of one of its gas layers, its spectra
continue to deviate significantly at low energy compared to other PCUs.
Therefore, although PCU \#\,0 was also on during the runs, we have not
included it in our spectral analysis.} data was available, whereas two and
even three detectors were on during significant fractions of time of the other
runs (PCUs \#\,1, 3 and 4). All the observations performed with PCA have been
reduced and analysed. A summary of the observations is shown in
Table~\ref{rxteobs}.

We have reduced the data using the HEAsoft FTOOLS package version 5.3.1. The
spectral analysis has been performed with Xspec 11.3.1. To carry out the data
reduction, we have created a filter file to avoid data taken with an elevation
angle with respect to the horizon smaller than 10 degrees, electron
contaminated data, data taken during the South Atlantic Anomaly passage and
data with an offset of the pointings bigger than 0.02 degrees. We have
accounted also for the detectors that were on during the observation. 

\begin{figure}
\plotone{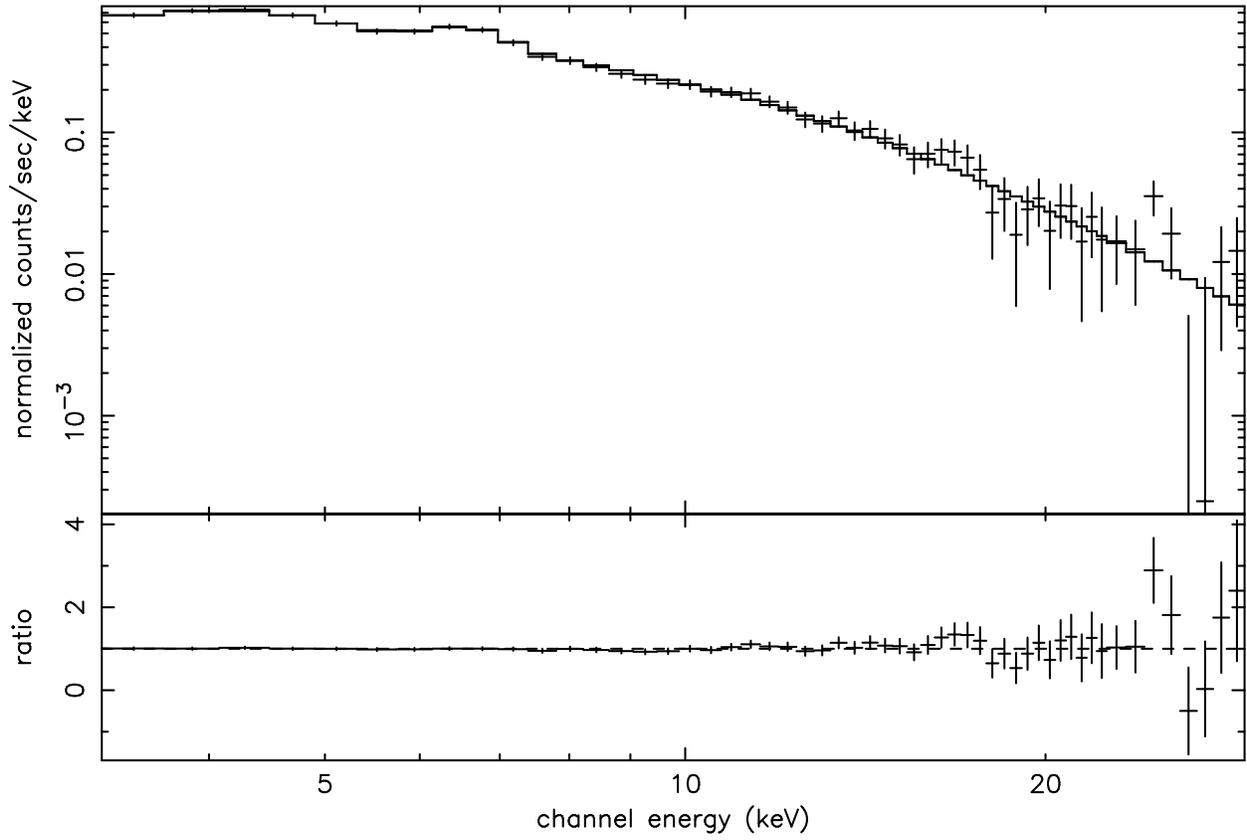}
\caption{X-ray spectrum of \lsff\ obtained with the data taken between phases 
0.78 and 0.84. The solid line represents the fit to the data. The lower panel 
shows the ratio between observed and fitted fluxes.}
\label{spec}
\end{figure}

For the 17 runs, we have tried to fit the spectra with an absorbed power-law
plus a Gaussian emission line and a disk black body components. For all of
them either the difference between models with and without a black body
component is insignificant or the black body flux is consistent with being
zero. Therefore, to estimate the photon index and the fluxes of both the
continuum and the line, we have fitted the data only with an absorbed
power-law plus a Gaussian line around 6.7~keV with fixed zero width (due to
the low spectral resolution of {\it RXTE}). Regarding absorption, we note that
inhomegeneities in the velocity and/or spatial distribution of the stellar
wind could produce a variable amount of hydrogen column density, as has been
observed in other wind-fed accreting systems like \object{GX~301$-$2}
\citep{white84}. Unfortunately, it is not possible to extract such kind of
information from our {\it RXTE} data of \ls\ due to the faint emission of the
source, the relatively low value of $N_{\rm H}$ and the low-energy threshold
of our data at 3~keV. Therefore, we have assumed a constant hydrogen column
density and fixed it to a value of $8.7\times 10^{21}$~cm$^{-2}$ (resulting
from the average between the one found by Reig et~al. (2003a) using the {\it
BeppoSAX} data and the one derived from the formula of Predehl \& Schmitt
(1995) applied to the quoted value for $A_V$ by McSwain et~al. (2004)). The
spectrum and the fit to the data taken at phase 0.78--0.84, during a
long-lasting maximum of emission, are presented in Fig.~\ref{spec}. The aspect
of the spectra and the fits obtained from the other observations are quite
similar to the one shown here. The date, time and orbital phase of the 17
runs, with their power-law indexes and unabsorbed fluxes for the power-law
component in the energy range 3--30~keV are presented in Table~\ref{rxteobs}. 

Due to the low source flux, the upper limits on the source variability within
a given run are not very constraining. The derived limits depend on a number
of factors, including the fitting functions used (power-laws, Lorentzians,
etc.) and the background subtraction (total background is about 30
cts~s$^{-1}$~PCU$^{-1}$ in the 3--30~keV energy range). Fast Fourier
transforms of the high time resolution data from all active PCUs were
performed to create power density spectra (PDS) in the 0.01--1000~Hz range.
Given the faint nature of the source, all available energy channels were
combined when making the PDS, and the obtained fractional variability
amplitude is generally less than 25\% (3$\sigma$). The PDS are dominated by
Poisson noise, and generally appear to have no significant noise components
which can be attributed to the source.

\subsection{Results} \label{res}

\begin{figure}
\plotone{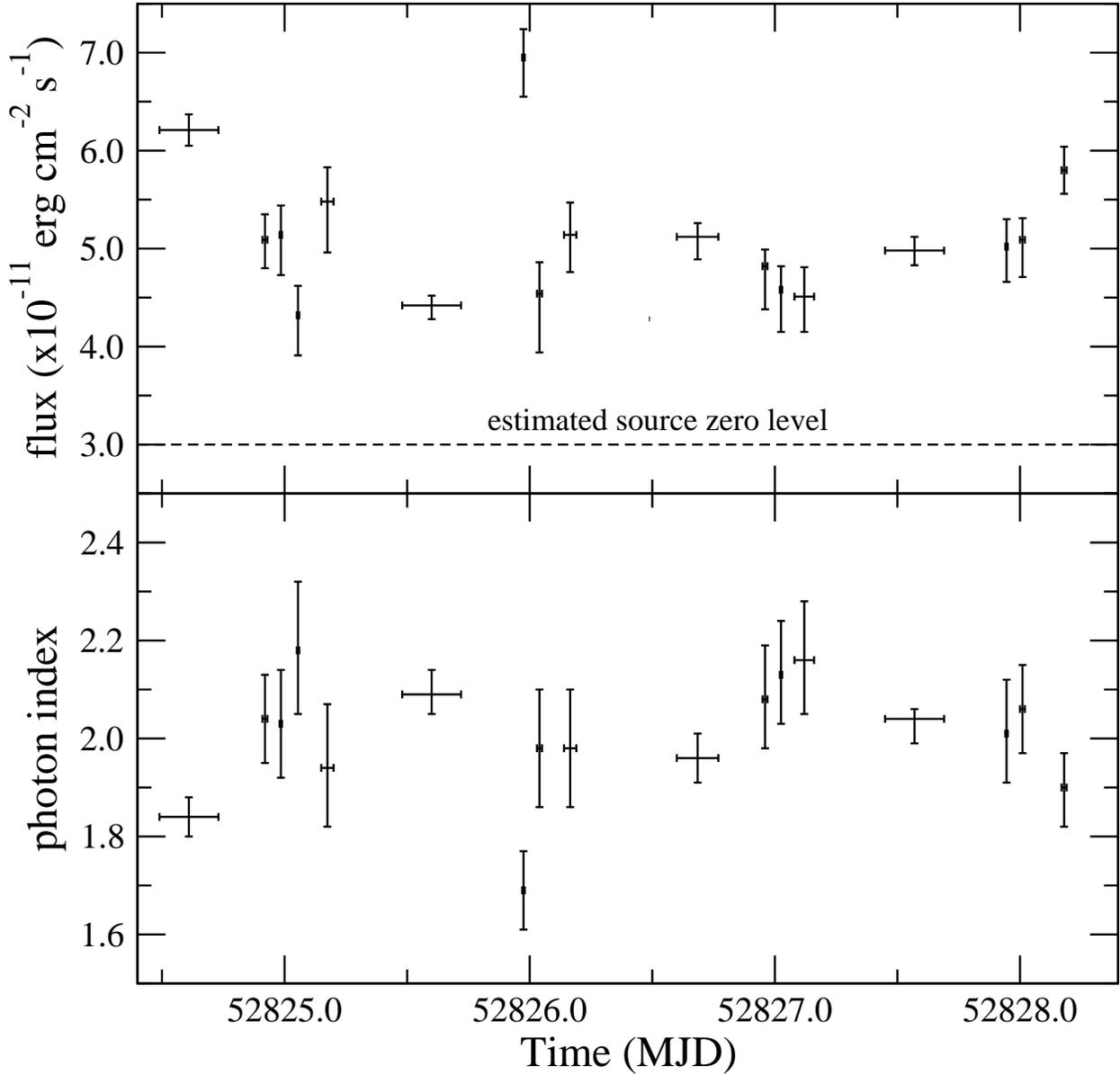}
\caption{Unabsorbed power-law flux in the energy range 3--30~keV (upper panel) 
and photon index (lower panel) along the observational period, without 
subtracting the diffuse background model. The vertical error bars show 
the 90\% of confidence, the horizontal error bars show the duration of 
the observation in time. The estimated source zero level is indicated 
by a dashed line in the upper panel.}
\label{flev1}
\end{figure}

\begin{figure}
\plotone{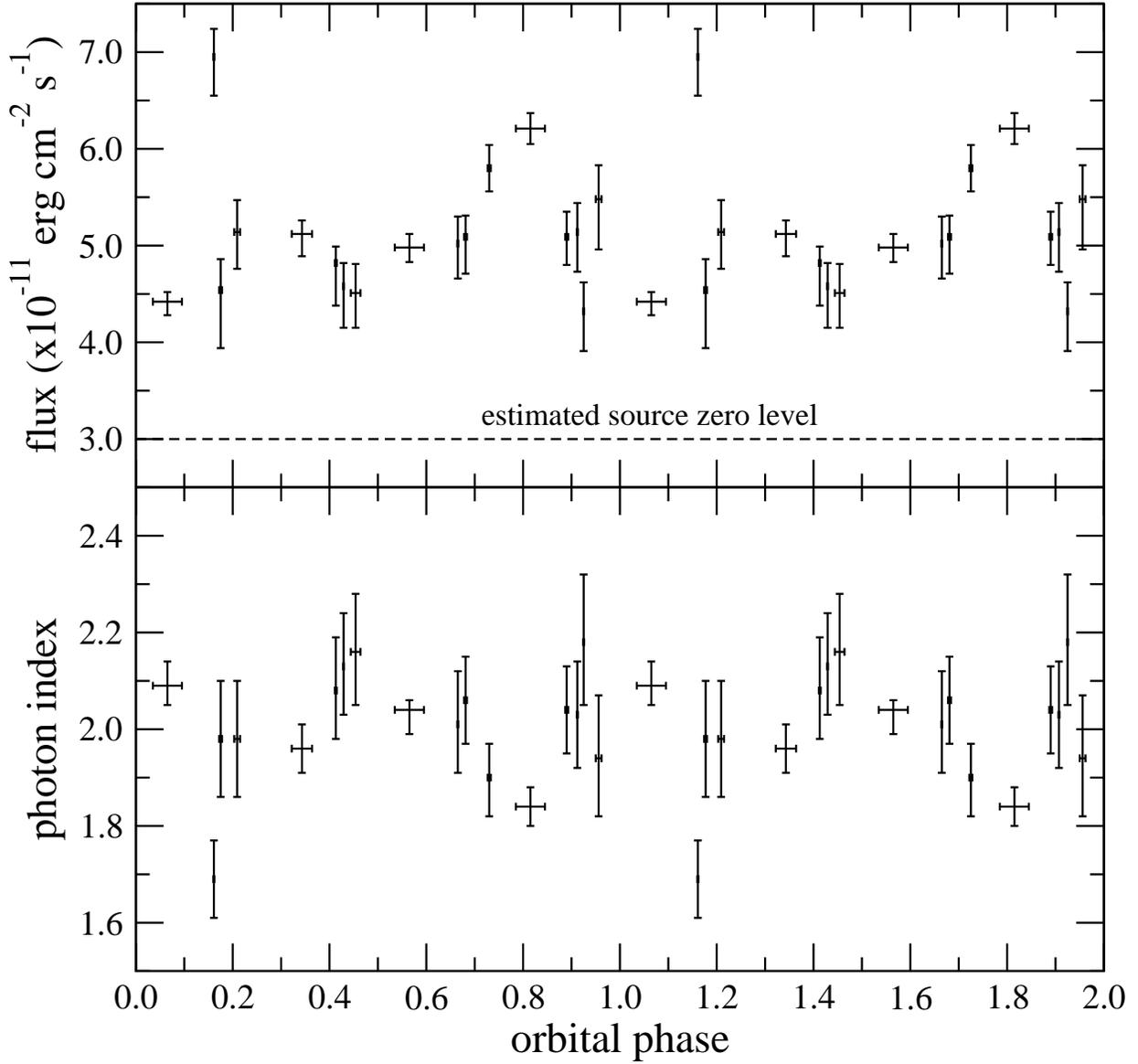}
\caption{Same as Fig.~\ref{flev1} but as a function of the orbital phase,
computed by using $P_{\rm orb}=3.9060 \pm 0.0002$~d and 
$t_0={\rm HJD}~2$,451,$943.09 \pm 0.10$. The periastron 
passage is at phase 0.0, and two orbital periods are shown 
for a better display.}
\label{flev2}
\end{figure}

The unabsorbed power-law flux and the photon index as a function of time are
presented in Fig.~\ref{flev1}. We find evidence of day to day variations in
both parameters. We show the same data folded with the orbital period in
Fig.~\ref{flev2}. It can be seen that the flux appears to be moderately
variable along the orbit, presenting a variation of a 50\% (see
Table~\ref{rxteobs}) with a long-lasting maximum at phase 0.8 and a
short-duration maximum around phase 0.16. This short-duration maximum, of
about half an hour, is more than three sigmas over the level of emission
indicated by the adjacent points, and it is not unique since other
short-duration peaks are present even within the longer runs. The plotted flux
in Fig.~\ref{flev2} has been corrected for the internal background. However,
this background subtracted flux still contains a contribution from the diffuse
background that is difficult to disentagle from the source spectrum and is
estimated to be about 2 counts per second per PCU\footnote{Although this value
has been found by the {\it RXTE} team after observations performed during the
2004 fall, we have assumed that during our observations, performed in 2003
July, it was roughly the same.}. Applying an absorbed Raymond plus power-law
model (see Valinia \& Marshall 1998) to this count rate in the 3--30~keV
energy interval, we obtain a diffuse background flux of about 
3$\times10^{-11}$~erg~cm$^{-2}$~s$^{-1}$. Thus, the ratio between the
long-lasting maximum and the minimum fluxes after subtracting the diffuse
background is $\sim$2.5.

\begin{figure}
\plotone{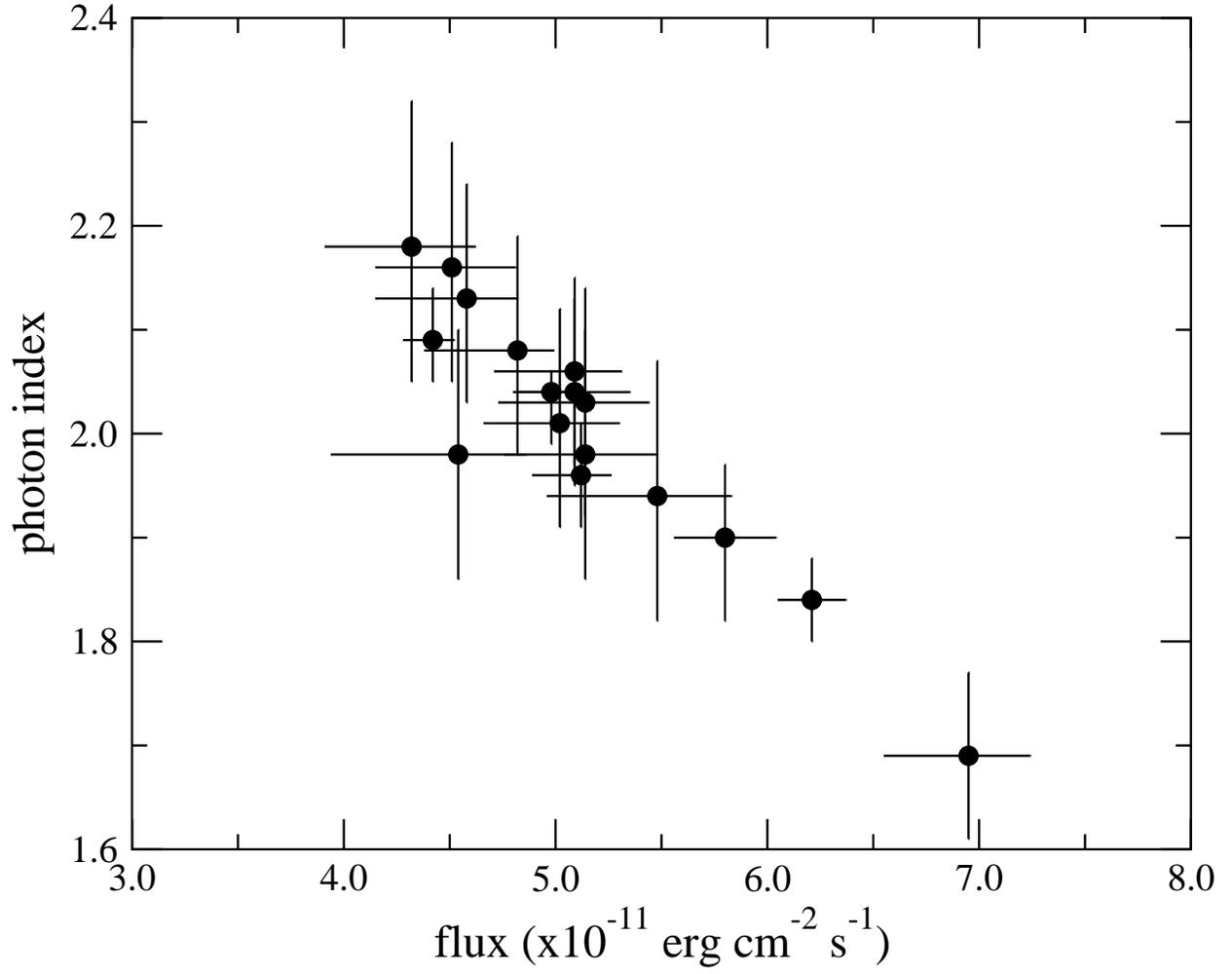}
\caption{Photon index as a function of unabsorbed power-law flux in the energy 
range 3--30~keV, without subtracting the diffuse background model. The errors 
are within the 90\% of confidence.}
\label{fluxg}
\end{figure}

In the lower panel of Fig.~\ref{flev2}, we show the evolution of the photon
index with the orbital phase. As can be seen, the photon index varies along
the orbital period, presenting a hardening close to periastron. Actually,
comparing in Fig.~\ref{flev2} the unabsorbed power-law flux and the photon
index evolutions at different phases of the orbit, an anti-correlation between
both appears. This is better illustrated in Fig.~\ref{fluxg}, where we plot
the photon index as a function of the unabsorbed power-law flux. It seems
that, when the flux is higher, the photon index is harder, and the contrary
happens when the flux is lower (this anti-correlation holds even within the
longer observations when split and analysed by parts). We stress here that
changes in flux and photon index cannot be due to changes in the hydrogen
column density for this weakly absorbed source, since we are working in the
energy range 3--30~keV, nor to diffuse background variations (internal
background ones are already taken into account) since, after applying the
diffuse background model, the anti-correlation is still there. We note that
the photon indexes calculated using this model are harder, in the range
$\sim$1.3--1.6, more in accordance with those obtained in previous X-ray
observations of the source, excluding the 1998 {\it RXTE} ones. Nevertheless,
the uncertainty in the diffuse background model determination at the \ls\
position requires to be cautious about these results. Moreover, the flux
variability does not seem to depend only on the photon index variability,
since when fixing it to an intermediate value of 1.9 the flux variation is
still significant. Another test has been done by fitting the unabsorbed flux
data given in Table~\ref{rxteobs} with a constant value. The obtained reduced
$\chi^2$ greater than 3 indicates that the fit is not good. Moreover, the
diffuse background can not introduce variability, since its value, estimated
by the {\it RXTE} team, remains constant within about a 10\% during periods of
time longer than the orbital period. Finally, no significant optical
variations have been observed in simultaneous photometric observations
\citep{marti04}, which indicates that the companion star showed a stable
behavior from the photometric point of view during our {\it RXTE}
observations, reinforcing the idea that the measured X-ray flux variability is
due to changes of the accretion rate in an eccentric orbit.

Although the data do not cover continuously the whole orbit, there appear to
be two types of variation timescales in the lightcurves. One apparently linked
to the orbital period of $\sim$3.9~days, the other of shorter duration, like
miniflares, of about 1~hour. No significant changes are detected through
simultaneous (although not strictly) spectroscopic observations of the
H$\alpha$ line along the orbit on day timescales \citep{casares05}. Thus,
stellar wind changes can be ruled out to explain the first type of
variability. In contrast, fast fluctuations in the stellar wind might be
related to the shorter timescale variability as significant variations on
timescales of minutes have been observed in radial velocity measurements of
\ls\ (see figure~2 of Casares et~al. (2005), around HJD~2,452,827.7). 
Unfortunately, there is no spectroscopic data right before
these miniflares observed in X-rays.

Our fits to the spectra are well consistent with the presence of a line at
6.7$\pm$0.1~keV in the data, as in the previous {\it RXTE} observations
\citep{ribo99}. However, as mentioned above, no other X-ray mission but {\it
RXTE} (which does not have imaging capabilities) reported the presence of the
iron line, so its origin could be in the Galactic plane regions behind the
source, and due to the so-called Galactic ridge emission. One possible test
for determining whether or not the line is related to \ls\ would be the
presence or not of line-flux variations correlated with continuum flux
changes. In fact, the different values of the line flux obtained on the
different runs are always consistent, at 90\%confidence, with their average
value of $\sim1.65\times10^{-4}$~photon~cm$^{-2}$~s$^{-1}$ per PCU beam of 0.9
square degree. For comparison, the value of the iron-line flux at the \ls\
position after measurement with the {\it Ginga} Large Area Proportional
Counters (LAC) is $\sim4.0\times10^{-4}$~photon~cm$^{-2}$~s$^{-1}$ per LAC
beam of 1\degr$\times$2\degr, with the large beam oriented perpendicular to
the Galactic plane (Yamauchi \& Koyama 1993). Therefore, this implies a flux
around $1.8\times10^{-4}$~photon~cm$^{-2}$~s$^{-1}$ per 0.9 square degree,
compatible with our individual measurements and slightly above the average
value we have obtained. Moreover, we have performed the analysis of the slew
data (when {\it RXTE} was pointing off the source) along the four days of
observation and found marginal evidence of the presence of a line (3$\sigma$)
with the flux similar to that obtained from the on-source data. Although slew
data have been difficult to model and we had to restrict the energy range to
3.0--10~keV, we have fitted first the data to a Bremsstrahlung obtaining a
reduced $\chi^2$ of 2, and afterwards to a Bremsstrahlung plus a Gaussian
line, improving the reduced $\chi^2$ value down to 1. The former facts seem to
rule out the possibility that the line comes from the source, although a deep
observation by an instrument with imaging capabilities, if possible around
periastron passage when the flux is higher, is still required to definitively
solve this question.

\section{Discussion} \label{disc}

\subsection{Orbital X-ray variability}

The {\it RXTE} observations presented here show evidence of flux variability
occurring on timescales of days which is anti-correlated to the photon index
of the spectrum. Since this happens while no changes are detected in the
companion star and the maximum X-ray flux occurs near periastron, we propose
that this variability is related to the orbital motion of the compact object
accreting from the companion star wind along an eccentric orbit. Actually the
emission peak seems to take place at phase 0.8, not right at periastron
passage. Leahy (2002) models a similar X-ray lightcurve observed in
\object{GX~301$-$2} through a non spherically symmetric stellar wind accreted
by the compact object following Bondi-Hoyle accretion.

We can consider that the wind of the companion star follows a $\beta$-law and
use a simple spherically symmetric Bondi-Hoyle accretion model (see Reig
et~al. 2003a for details) to study the ratio of maximum to minimum flux, to be
compared to the observed one. For this purpose we have assumed the following
parameters: $M_{\rm opt}=40~M_\odot$, $R_{\rm opt}=10~R_\odot$, $\dot{M}_{\rm
opt}=1 \times 10^{-7}~M_\odot$~yr$^{-1}$ (average between the value given in
Kudritzki \& Puls 2000 from the wind-momentum relationship and the value
inferred from optical spectroscopy by McSwain et~al. 2004), $v_{\infty} =
2440$~km~s$^{-1}$ (from observations by McSwain et~al. 2004), $\beta=0.8$, and
finally a neutron star of $M_{\rm X}=1.4~M_\odot$ with a radius of $R_{\rm
X}=15$~km in an eccentric orbit with $P_{\rm orb} = 3.9060$~days and $e=0.35$.
(Casares et~al. 2005). These parameters provide an expected variability with a
ratio of $\sim$10 between the maximum and minimum flux. Even when considering
the uncertainties in all the involved parameters, ratios between 8--17 are
found. These ratios are much lower than the variability ratio $\sim$40
considering the previous eccentricity value of about 0.5 \citep{mcswain04},
but still too large to explain the ratio of $\sim$2.5 observed by {\it RXTE}.

A possible solution that would easily reconcile the theoretical and the
observed flux ratio is the presence of an accretion disk around the compact
object. This would smoothen the accretion rate of the compact object, implying
lower and longer emission bumps around periastron passage in the lightcurves.
Therefore, the difference between the expected and the observed flux ratio
could be the first hint, in the X-ray domain and independent of the presence
of relativistic jets, of the existence of an accretion disk in \ls, although
it is not clear what kind of disk can be present at such a low ratio of X-ray
to Eddington luminosity. It should be noted that for the short-timescale
variations or miniflares, although they present the same flux/photon index
anti-correlation as the rest of the observations suggesting to be intrinsic to
the source, the disk smoothing might not operate in the same manner as for
long-lasting peaks, since it is the wind itself that changes rapidly. This
could affect the accretion process in a more violent manner (e.g. inducing
unsteady accretion and subsequent sudden X-ray flux increase) than a slow
changing flow, as it would be the case for an approximately constant mass-loss
rate crossed by the accreting compact object.

{\it BeppoSAX} observations show a decrease in flux of about a 50\% between
the periastron passage at phase 0.0 and the end of the observations at phase
0.21 (figure~1 of Reig et~al. 2003a). Interestingly, this is not too far from
the flux reduction in this phase interval predicted by the accretion model
discussed above, and we remind the reader that {\it BeppoSAX} had imaging
capabilities, preventing the influence of background sources in the data. We
note that the position of the maximum of emission in the {\it BeppoSAX}
observations is consistent to be before periastron, like in the observations
presented here. Therefore, these observations are complementary to the {\it
RXTE} ones reported here and give clear support to the orbital variability as
due to the eccentric orbit.

Concerning the long-term correlations in the X-ray emission and the stellar mass-loss
rate, the new values for the
equivalent width of the H$\alpha$ absorption line of $\sim$2.8~\AA\ \citep{casares05}
are similar to those obtained during the epochs when {\it ASCA} and {\it Chandra}
observed the source. Moreover, taking into account the diffuse background emission, the
average X-ray flux truly coming from the source during our observations is about
2$\times10^{-11}$~erg~cm$^{-2}$~s$^{-1}$, similar also to the fluxes observed during
{\it ASCA} and {\it Chandra} observations. Therefore, the apparent correlation between
changes in the mass-loss rate of the stellar companion and the X-ray emission,
suggested by Reig et~al. (2003a) and  supported by McSwain et~al. (2004), seems to be
confirmed by these new spectoscopic and X-ray results.

\subsection{Proposed origin for the X-ray emission}

The signature of the existence of an accretion disk around the compact object
is not present in any of the obtained X-ray spectra of the source and no
cut-off is present in data up to 30~keV. Actually, the possible detection of
\ls\ by instruments working in the energy range beyond the {\it RXTE} one
(BATSE, Harmon et~al. \citep{harmon04}; COMPTEL, Strong et~al. 2001; EGRET,
Paredes et~al. 2000) seems to preclude the existence of any cut-off at
energies beyond the {\it RXTE} range, typical of the thermal Comptonization
spectra for emission produced in a corona. Therefore, we propose that the
X-ray emission is due to inverse Compton (IC) and/or synchrotron processes
within the relativistic jets of \ls. In our scenario, an increase in the
accretion rate would imply a higher electron acceleration efficiency in the
jets, filling with fresh particles the higher energy range of the electron
energy distribution. This would harden the X-ray spectrum, which is actually
observed by {\it RXTE} through the flux/photon index anti-correlation. The
smooth evolution of the flux could be explained mainly, as it is mentioned
above, by the dynamical timescales of a faint underlying accretion disk, which
would smoothen the emission variation around the periastron passage. It has
been considered by several authors that relativistic electrons in the jet of a
microquasar could produce significant X-ray emission by IC upscattering of the
stellar photons or their own synchrotron photons (see Kaufman Bernad\'o,
Romero, \& Mirabel 2002; Georganopoulos, Aharonian, \& Kirk 2002; Reig,
Kylafis, \& Giannios 2003b; Bosch-Ramon, Romero, \& Paredes 2005). Optically
thin synchrotron radiation from a jet could also dominate at X-rays, depending
on the matter density and magnetic field conditions within the jet (Markoff,
Falcke, \& Fender 2001). Jet dominated states could be common and the disk
radiation would appear covered by the non-thermal jet emission (Fender, Gallo,
\& Jonker 2003). These jet dominated states are associated with the low-hard
state of black hole X-ray binaries. The nature of the compact object in \ls\
is still unknown (although in could be a black-hole, see Casares et~al. 2005)
and its relatively high radio flux density and optically thin radio
spectrum do not agree with the observed values found in black holes (Gallo,
Fender, \& Pooley 2003). However, the observed X-ray spectrum is always very
similar to the ones of black holes in the low-hard state: a power-law with a
photon index around 1.5 (see Tables~\ref{priorobs} and \ref{rxteobs}).

\section{Summary} \label{sum}

We have reported recent {\it RXTE} observations of \ls\ covering a full
orbital period of 4~days. The obtained results show moderate day to day
variability that is interpreted as due to changes in the accretion rate
because of the orbital motion of the compact object along an eccentric orbit,
with an increase of emission right before periastron passage. In addition, the
observed short-timescale flux variations (miniflares) are likely related to
observed fast variations in the stellar wind properties. These observational
results give support to the accretion scenario, making other possibilities
like the young non-accreting pulsar scenario (Martocchia et al. 2005) to be
more unlikely. In parallel with the power-law flux variability, the flux of
the iron line, detected at $\sim$6.7~keV like in previous {\it RXTE}
observations, has remained constant within the errors and similar to the
expected background line flux. Furthermore, slew data of these {\it RXTE}
observations present marginal evidences of the iron line as well, and the line
has not been detected by X-ray satellites with imaging capabilities. All these
facts make us suggest it is a background feature. No disk signatures are found
in the spectra, although this is not rare for other X-ray binaries in the
low-hard state. In fact, the presence of an accretion disk could explain the
apparent smoothening of the flux changes along the orbit observed by {\it
RXTE} with respect to what is predicted by using a Bondi-Hoyle accretion
model. Finally, there appears to be an anti-correlation between the flux and
the X-ray photon index along the orbit, reaching a maximum of emission and the
hardest spectrum around phase 0.8. We suggest that a scenario where the
X-rays come from IC interactions and/or synchrotron emission within a
relativistic jet could explain this anti-correlation, as well as the absence
of cut-off at hard X-rays. Although further theoretical work must be done, new
observations to constrain the X-ray flux variability between periastron and
apastron passages are necessary to better understand the orbital variability
of \ls.

\acknowledgments

We thank the referee Jean Swank for relevant and valuable comments and
suggestions that helped to improve the paper significantly.
This research has made use of data obtained by the {\it RXTE} satellite. We
are grateful to J. Casares and collaborators for allowing us to publish 
the new orbital ephemeris prior to their publication, and for providing the 
equivalent width of H$\alpha$ in contemporaneous spectroscopic observations.
V.B-R., J.M.P., M.R. and J.M. acknowledge partial support by DGI of the
spanish Ministerio de Educaci\'on y Ciencia (former Ministerio de Ciencia y
Tecnolog\'{\i}a) under grants AYA-2001-3092, AYA2004-07171-C02-01, and
AYA2004-07171-C02-02, as well as additional support from the European Regional
Development Fund (ERDF/FEDER).
During this work, V.B-R. has been supported by the DGI of the spanish Ministerio de Educaci\'on y Ciencia under the fellowship FP-2001-2699. 
M.R. acknowledges support by a Marie Curie Fellowship of the European
Community programme Improving Human Potential under contract number
HPMF-CT-2002-02053.

\clearpage

\begin{deluxetable}{ccccc}
\tablecaption{Summary of previous X-ray observations. 
\label{priorobs}}
\tablewidth{0pt}
\tablehead{
 & & & & \colhead{Extrapolated 3--30~keV Flux} \\
\colhead{Date} & \colhead{Mission} & \colhead{Phase} & \colhead{Photon Index} & \colhead{($\times10^{-11}$~erg~cm$^{-2}$~s$^{-1}$)}}
\startdata
1998 Feb 08 & {\it RXTE}     & 0.803--0.837 & 1.95$\pm$0.02        & 5.6 \\
1998 Feb 08 & {\it RXTE}     & 0.006--0.057 & 1.95$\pm$0.02        & 4.6 \\
1998 Feb 16 & {\it RXTE}     & 0.039--0.059 & 1.95$\pm$0.02        & 4.6 \\
1999 Oct 04 & {\it ASCA}     & 0.382--0.471 & $\sim$1.6 ~~~~~      & 2.7 \\
2000 Oct 08 & {\it BeppoSAX} & 0.969--0.205 & 1.8$\pm$0.2 ~~       & 0.8 \\
2002 Sep 10 & {\it Chandra}  & 0.690--0.720 & 1.14$\pm$0.20        & 2.3 \\
2003 Mar 08 & {\it XMM}  & 0.526--0.557 & 1.56$^{+0.02}_{-0.05}$ ~ & 2.0 \\
2003 Mar 27 & {\it XMM}  & 0.533--0.564 & 1.49$^{+0.05}_{-0.04}$ ~ & 2.0 \\
\enddata
\tablecomments{Errors are within the 90\% of confidence.}
\end{deluxetable}

\clearpage

\begin{deluxetable}{ccccc}
\tablecaption{Summary of {\it RXTE} observations. 
\label{rxteobs}}
\tablewidth{0pt}
\tablehead{
\colhead{Middle Time} & \colhead{Duration} & & & \colhead{Unabsorbed 3--30~keV Flux } \\
\colhead{(MJD)} & \colhead{(day)} & \colhead{Phase} & \colhead{Photon Index} & \colhead{($\times10^{-11}$~erg~cm$^{-2}$~s$^{-1}$)}}
\startdata
52824.61 & 0.24 & 0.782--0.842 & $1.84^{+0.04}_{-0.04}$ & $6.21^{+0.16}_{-0.16}$ \\
52824.92 & 0.02 & 0.888--0.894 & $2.04^{+0.09}_{-0.09}$ & $5.09^{+0.26}_{-0.29}$ \\
52824.99 & 0.01 & 0.907--0.910 & $2.03^{+0.11}_{-0.11}$ & $5.14^{+0.30}_{-0.41}$ \\
52825.06 & 0.01 & 0.924--0.926 & $2.18^{+0.14}_{-0.13}$ & $4.32^{+0.30}_{-0.41}$ \\
52825.18 & 0.05 & 0.950--0.963 & $1.94^{+0.13}_{-0.12}$ & $5.48^{+0.35}_{-0.52}$ \\
52825.60 & 0.24 & 0.034--0.095 & $2.09^{+0.05}_{-0.04}$ & $4.42^{+0.10}_{-0.14}$ \\
52825.98 & 0.01 & 0.159--0.162 & $1.69^{+0.08}_{-0.08}$ & $6.95^{+0.29}_{-0.40}$ \\
52826.04 & 0.02 & 0.176--0.181 & $1.98^{+0.12}_{-0.12}$ & $4.54^{+0.32}_{-0.60}$ \\
52826.17 & 0.05 & 0.202--0.217 & $1.98^{+0.12}_{-0.12}$ & $5.14^{+0.33}_{-0.38}$ \\
52826.69 & 0.17 & 0.321--0.364 & $1.96^{+0.05}_{-0.05}$ & $5.12^{+0.14}_{-0.23}$ \\
52826.96 & 0.02 & 0.411--0.415 & $2.08^{+0.11}_{-0.10}$ & $4.82^{+0.17}_{-0.44}$ \\
52827.03 & 0.01 & 0.429--0.431 & $2.13^{+0.11}_{-0.10}$ & $4.58^{+0.24}_{-0.43}$ \\
52827.12 & 0.08 & 0.445--0.465 & $2.16^{+0.12}_{-0.11}$ & $4.51^{+0.30}_{-0.36}$ \\
52827.57 & 0.24 & 0.539--0.600 & $2.04^{+0.02}_{-0.05}$ & $4.98^{+0.14}_{-0.15}$ \\
52827.95 & 0.01 & 0.663--0.667 & $2.01^{+0.11}_{-0.10}$ & $5.02^{+0.28}_{-0.36}$ \\
52828.01 & 0.02 & 0.680--0.685 & $2.06^{+0.09}_{-0.09}$ & $5.09^{+0.22}_{-0.38}$ \\
52828.18 & 0.02 & 0.724--0.728 & $1.90^{+0.07}_{-0.08}$ & $5.80^{+0.24}_{-0.24}$ \\
\enddata
\tablecomments{Errors are within the 90\% of confidence.}
\end{deluxetable}

\clearpage

\clearpage 

\clearpage

\end{document}